\documentclass{elsart}

\usepackage{graphicx}

\usepackage{amsmath,amssymb}
\usepackage[sort&compress,numbers,square]{natbib}
\begin{document}

\begin{frontmatter}



\title{Performance of the neutron polarimeter NPOL3
for high resolution measurements}


\author[Kyushu]{T.~Wakasa}
\author[Kyushu]{Y.~Hagihara}
\author[Tokyo]{M.~Sasano}
\author[Kyushu]{S.~Asaji}
\author[RCNP]{K.~Fujita}
\author[RCNP]{K.~Hatanaka}
\author[Kyushu]{T.~Ishida}
\author[CNS]{T.~Kawabata}
\author[Tokyo]{H.~Kuboki}
\author[CNS]{Y.~Maeda}
\author[Kyushu]{T.~Noro}
\author[Tokyo]{T.~Saito}
\author[Tokyo]{H.~Sakai}
\author[RCNP]{Y.~Sakemi}
\author[RIKEN]{K.~Sekiguchi}
\author[RCNP]{Y.~Shimizu}
\author[RCNP]{A.~Tamii}
\author[RCNP]{Y.~Tameshige}
\author[Tokyo]{K.~Yako}

\address[Kyushu]{Department of Physics, Kyushu University, 
Higashi, Fukuoka 812-8581, Japan}
\address[RCNP]{Research Center for Nuclear Physics, Osaka University,
Ibaraki, Osaka 567-0047, Japan}
\address[CNS]{Center for Nuclear Study, The University of Tokyo,
Bunkyo, Tokyo 113-0033, Japan}
\address[Tokyo]{Department of Physics, The University of Tokyo,
Bunkyo, Tokyo 113-0033, Japan}
\address[RIKEN]{The Institute of Physical and Chemical Research,
Wako, Saitama 351-0198, Japan}

\begin{abstract}
 We describe the neutron polarimeter NPOL3 for the measurement 
of polarization transfer observables $D_{ij}$ with a typical 
high resolution of $\sim$300 keV at $T_n$ $\simeq$ 200 MeV.
 The NPOL3 system consists of three planes of neutron detectors.
 The first two planes for neutron polarization analysis 
are made of 20 sets of one-dimensional position-sensitive 
plastic scintillation counters with a size of 
100 cm $\times$ 10 cm $\times$ 5 cm, and they cover the 
area of 100 $\times$ 100 $\mathrm{cm}^2$.
 The last plane for detecting doubly scattered neutrons or 
recoiled protons is made of the two-dimensional 
position-sensitive liquid scintillation counter 
with a size of 100 cm $\times$ 100 cm $\times$ 10 cm.
 The effective analyzing powers $A_{y;\mathrm{eff}}$ 
and double scattering efficiencies $\epsilon_{\mathrm{D.S.}}$
were measured
by using the three kinds of polarized neutrons from the 
${}^{2}{\rm H}(\vec{p},\vec{n})pp$, 
${}^{6}{\rm Li}(\vec{p},\vec{n}){}^{6}{\rm Be}(\mathrm{g.s.})$, and
${}^{12}{\rm C}(\vec{p},\vec{n}){}^{12}{\rm N}(\mathrm{g.s.})$ reactions 
at $T_p$ = 198 MeV.
 The performance of NPOL3 defined as 
$\epsilon_{\mathrm{D.S.}}(A_{y;\mathrm{eff}})^2$ are 
similar to that of the Indiana Neutron POLarimeter (INPOL)
by taking into account for the counter configuration difference 
between these two neutron polarimeters.
\end{abstract}

\begin{keyword}
Neutron polarimeter \sep position sensitive detector \sep
polarization transfer observables \sep time of flight
\PACS 29.30.Hs \sep 29.40.Mc \sep 28.20.Cz \sep 25.40.Kv
\end{keyword}
\end{frontmatter}

\section{Introduction}
\label{sec:intro}

 The polarization transfer observables $D_{ij}$ for the charge-exchange
$(\vec{p},\vec{n})$ reaction at intermediate energies $T_p$ $>$ 100 MeV
provides a potentially rich source of information not only on 
nuclear responses but also on effective interactions, 
and their extensive studies have been performed at the Los Alamos 
Meson Physics Facility (LAPMF), the Indiana University Cyclotron 
Facility (IUCF), and the Research Center for Nuclear Physics (RCNP).
 An example of these studies is the search for a pionic enhancement
in nuclei via  measurements of 
a complete set of $D_{ij}$ for $(\vec{p},\vec{n})$ quasielastic 
scatterings (QES).
 The spin-longitudinal (pionic) response function $R_L$ is theoretically
expected to be enhanced relative to the spin-transverse response 
function $R_T$ \cite{plb_92_153_1980,npa_379_429_1982}.
 The enhancement of $R_L$
is attributed to the collectivity induced by the attraction of the one-pion 
exchange interaction, and has aroused much interest 
in connection with both the precursor phenomena of the pion
condensation \cite{plb_76_182_1978,prl_42_1034_1979,%
plb_89_327_1980,plb_92_153_1980,npa_379_429_1982} 
and the pion excess in the nucleus 
\cite{plb_128_107_1983,plb_128_112_1983,prl_51_763_1983,%
prd_29_398_1984,prc_57_1210_1998}.
 Surprisingly, the experimentally extracted $R_L/R_T$ ratios 
are less than or equal to unity \cite{prl_73_3516_1994,prc_59_3177_1999}, 
which contradicts the theoretical predictions 
of the enhanced $R_L$ and the quenched $R_T$.
 Recent analyses of the QES data 
\cite{prc_69_054609_2004,nucl_ex_0411055} 
show the pionic enhancement in the spin-longitudinal 
cross section which well represents the $R_L$.
 The discrepancy in $R_L/R_T$ might be due to the 
effects of the medium modifications of the 
effective NN interaction.
 These effects could be studied by measuring a complete set
of $D_{ij}$ for stretched states \cite{prl_78_1636_1997,rcnp_e236},  
which requires a relatively better energy resolution of 
$\sim$500 keV compared with the energy resolution of 
2--3 MeV in the QES measurements.

 Another requirement for high resolution is for the measurement of 
the Gamow-Teller (GT) unit cross section $\hat{\sigma}_{\rm GT}$.
 Recently Yako et al. \cite{nucl_ex_0411011} applied 
multipole decomposition analysis  both to their 
${}^{90}{\rm Zr}(n,p)$ data and to the ${}^{90}{\rm Zr}(p,n)$ data 
in Refs.~\cite{prc_55_2909_1997,jpsj_73_1611_2004}, 
and obtained the GT quenching factor $Q$ = 0.88 $\pm$ 0.03 $\pm$ 
0.16.
 The first uncertainty contains the uncertainties both
of the MDA and 
of the estimation of the isovector spin-monopole (IVSM) contributions,
and the second uncertainty originates from the uncertainty of 
$\hat{\sigma}_{\rm GT}$.
 This large $Q$ value clearly indicates that the configuration 
mixing is the main 
mechanism of the quenching and thus the admixture of the $\Delta$h 
states into the low-lying states plays a minor role. 
 However, a relatively large uncertainty of $\hat{\sigma}_{\rm GT}$ 
makes it difficult to draw a definite conclusion from this 
$Q$ value.
 In principle, a precise  $\hat{\sigma}_{\rm GT}$ value 
can be obtained by measuring 
the cross section at $0^{\circ}$ for the ground or low-lying 
discrete GT state whose $B({\rm GT})$ value is known by the 
$\beta$-decay \cite{rcnp_e218}.
 In practice, such a measurement is hampered by a poor 
energy resolution of the neutron time-of-flight (TOF) system.
 Actually, for the TOF facility at RCNP, the energy resolution is
limited to be $\sim$1.6 MeV at $T_p$ $\simeq$ 300 MeV 
due to the thick neutron counter thickness of NPOL2 
\cite{nima_404_355_1998} compared to the relatively short 
flight pass length 100 m \cite{nima_369_120_1996}. 
 This poor energy resolution does not allow such direct 
determination of $\hat{\sigma}_{\rm GT}$. 

 The demand for higher resolution necessary for nuclear 
spectroscopy led to the design and construction of the 
new neutron detector and polarimeter NPOL3.
 The neutron detector should be designed so that 
the final energy resolution is better than $\sim$500 keV 
in order to resolve GT and stretched states 
from their neighboring states.
 This can be achieved by using the neutron detector material 
thinner than the NPOL2.
 Furthermore, the NPOL3 is designed to realize the $D_{ij}$ 
measurement which is important to determine the effective 
NN interaction.

 In Section~\ref{sec:imp}, we present the results of the 
simulation performed to design the NPOL3 system.
 In Section~\ref{sec:single}, we will describe the NPOL3 
and its performance for the neutron detector.
 Sections~\ref{sec:calib} and \ref{sec:opt} are devoted 
to the calibration and optimization 
procedures of the NPOL3 for the neutron polarimeter.
 In Sections~\ref{sec:double} and \ref{sec:discuss}, 
we will discuss the results of 
the calibration.
 A summary is given in Section~\ref{sec:summary}.

\section{Improvements of time and energy resolution}
\label{sec:imp}

\subsection{Time and energy resolution}

 At intermediate energies where neutron kinetic 
energies $T_n$ are determined by the TOF
technique over a long flight path length of 
$\sim$100 m, a large volume of the scintillator 
is required in order to achieve a sufficient neutron 
detection efficiency.
 The energy resolution $\Delta T_n$ by the TOF technique 
is related to the uncertainties both of 
timing $\Delta t/t$ and flight path length $\Delta L/L$ 
as 
\begin{equation}
\frac{\Delta T_n}{T_n} = 
\gamma(\gamma+1)
\sqrt{
 \left(\frac{\Delta t}{t}\right)^2
+\left(\frac{\Delta L}{L}\right)^2
},
\label{eq:DeltaTn}
\end{equation}
with
\begin{equation}
\gamma = 1 + \frac{T_n}{m_n},
\end{equation}
where $m_n$ is the neutron mass.
 $\Delta t$ is the uncertainty of the flight time $t$ 
which originates from the time spread of the incident 
proton beam and the time resolution of the neutron counter.
 $\Delta L$ is the uncertainty of the flight path length $L$ 
which comes from the thickness of the counter.
 The charged particle scattered by an incident 
neutron passes the neutron counter and optical photons are 
generated along the path.
 The number of photons $N_{\rm photon}$ 
is proportional to the thickness $\Delta L$ 
and the time resolution $\Delta t$ becomes better by 
increasing $N_{\rm photon}$.
 However, the thickness $\Delta L$ is directly related to 
the energy resolution $\Delta T_n$ as is seen in 
Eq.~(\ref{eq:DeltaTn}).
 Thus the $\Delta L$ dependence on $\Delta T_n$ is rather 
complicated since $\Delta t$ is also the function of $\Delta L$.

\subsection{Monte-Carlo simulation with GEANT4}

\begin{figure}
\begin{center}
\includegraphics[width=0.9\linewidth,clip]{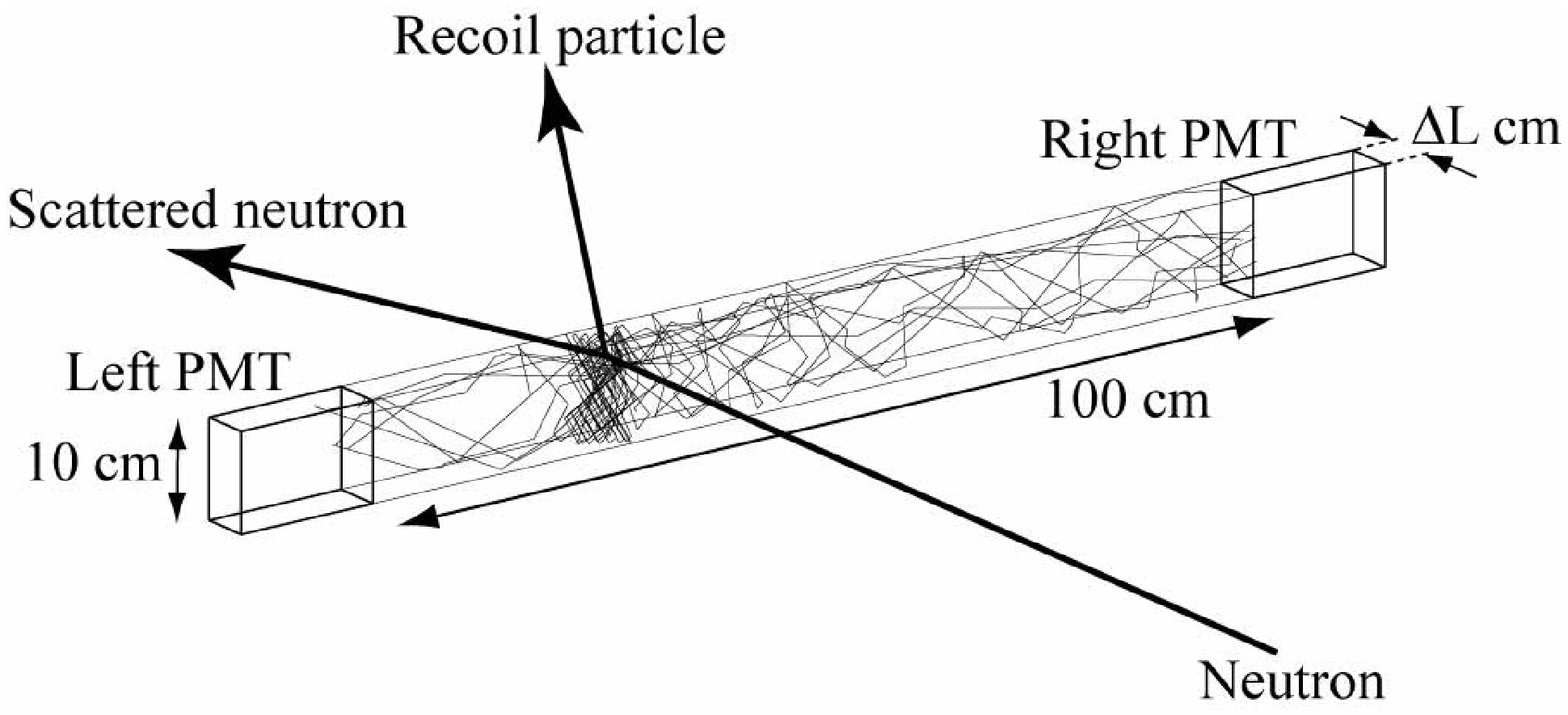}%
\end{center}
\caption{
 Illustration of the Monte-Carlo simulation with {\sc geant4}.
 The neutron beam with $T_n$ = 200 or 300 MeV bombard a 
plastic scintillator with a size of 100 cm $\times$ 10 cm 
$\times$ $\Delta L$ cm.
 The optical photons generated by recoiled charge particles 
propagate in the scintillator and are detected by PMTs at both
ends.
 The time information, $t_L$ and $t_R$, deduced from PMT 
signals are used to evaluate the time of flight $t$ and 
its width $\Delta t$.
\label{fig:hodo}}
\end{figure}

\begin{figure}
\begin{center}
\includegraphics[width=0.7\linewidth,clip]{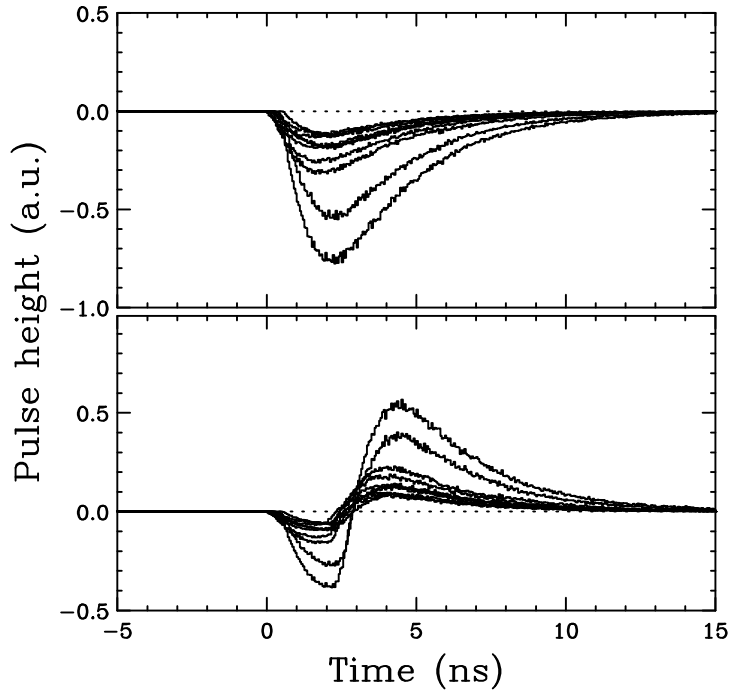}%
\end{center}
\caption{
 The pulse-height (upper panel) and CFD timing (lower panel) 
spectra generated by the Monte-Carlo simulation with {\sc geant4}.
 See text for details.
\label{fig:cfd}}
\end{figure}

 We performed the Monte-Carlo simulation in order to describe the 
energy resolution $T_n$ quantitatively as a function of $\Delta L$.
 The computer library {\sc geant4} \cite{geant4} 
was used to simulate the 
neutron-induced reactions in scintillator material, the generation 
of optical photons, and their propagation.
 The scintillator material is assumed to be the plastic scintillator 
Saint-Gobain \cite{url_saint_gobain} 
Bicron BC408 with a hydrogen to carbon ratio of H/C = 1.1.
 The size of the scintillator is 100 cm $\times$ 10 cm $\times$ 
$\Delta L$ cm, and its configuration is the one-dimensional 
position-sensitive detector (hodoscope) 
with a thickness of $\Delta L$ cm (see Fig.~\ref{fig:hodo}).
 The length of 100 cm is same as that of the 
NPOL2, and the width is fixed to be 10 cm.
 At both ends (L:left and R:right) of the scintillator, 
the timing information of arrived optical photons is accumulated.
 The neutron flight time $t$ can be deduced by using 
timing information, $t_L$ and $t_R$, at both ends as 
\begin{equation}
t = \frac{t_L+t_R}{2}-t_{\mathrm{RF}},
\end{equation}
where $t_{\rm RF}$ corresponds to the timing information 
for an accelerator RF signal in the experiment.
 In practice, photo-multiplier tubes (PMTs) are mounted at both 
ends and optical photons are converted to electric signals.
 Each anode signal from PMT is fed to a constant fraction 
discriminator (CFD) to create a fast logic timing signal.
 We simulated the operation of the CFD.
 The upper panel of Fig.~\ref{fig:cfd} is the pulse-height 
spectrum generated by our simulation, and the lower panel 
is the simulated CFD timing spectrum.
 The timing information is determined 
from the zero-crossing point of the CFD spectrum.
 Thus the resulting $t_L$ and $t_R$ are almost independent 
of an arbitrarily set discrimination level of the CFD.
 The time spread of the zero-crossing points in Fig.~\ref{fig:cfd}
corresponds to the counter-thickness ($\Delta L$) effect
to $\Delta t$.

\begin{figure}
\begin{center}
\includegraphics[width=0.7\linewidth,clip]{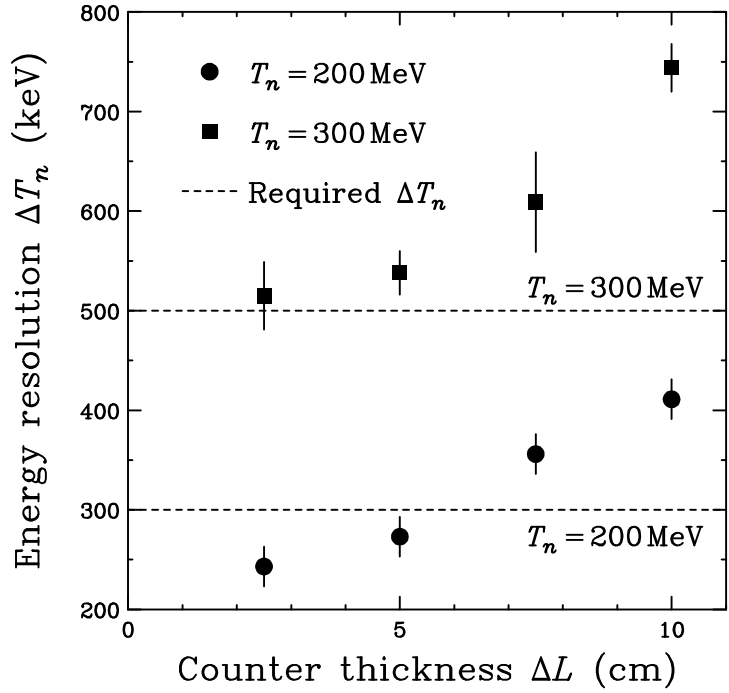}%
\end{center}
\caption{
 The simulation results for the energy resolution $\Delta T_n$
as a function of the counter thickness $\Delta L$.
 The filled circles and the filled boxes represent the 
results for the measurement at $T_p$ = 200 and 300 MeV, respectively, 
as explained in the text.
\label{fig:deltatn}}
\end{figure}

 Figure~\ref{fig:deltatn} represents the expected energy resolutions 
$\Delta T_n$ evaluated from the simulation results 
as a function of $\Delta L$.
 The filled circles and the filled boxes correspond to the 
results for the measurement at $T_n$ = 200 and 300 MeV, respectively.
 In the simulation, the flight path length $L$ is 100 m and 
the time spread of the proton beam is assumed 
to be 200 ps FWHM which is a typical value of the beam 
from the RCNP Ring cyclotron.
 The $\Delta T_n$ values clearly depend on $\Delta L$ 
for both $T_n$ = 200 and 300 MeV.
 It is found that the counter thickness $\Delta L$ should be 
thinner than 5 cm in order to achieve the required final 
resolution of $\sim$500 keV at $T_n$ = 300 MeV.
 Thus we have employed the hodoscope counters with a thickness 
of 5cm in the NPOL3.

\subsection{Comparison with experimental results}

\begin{figure}
\begin{center}
\includegraphics[width=0.7\linewidth,clip]{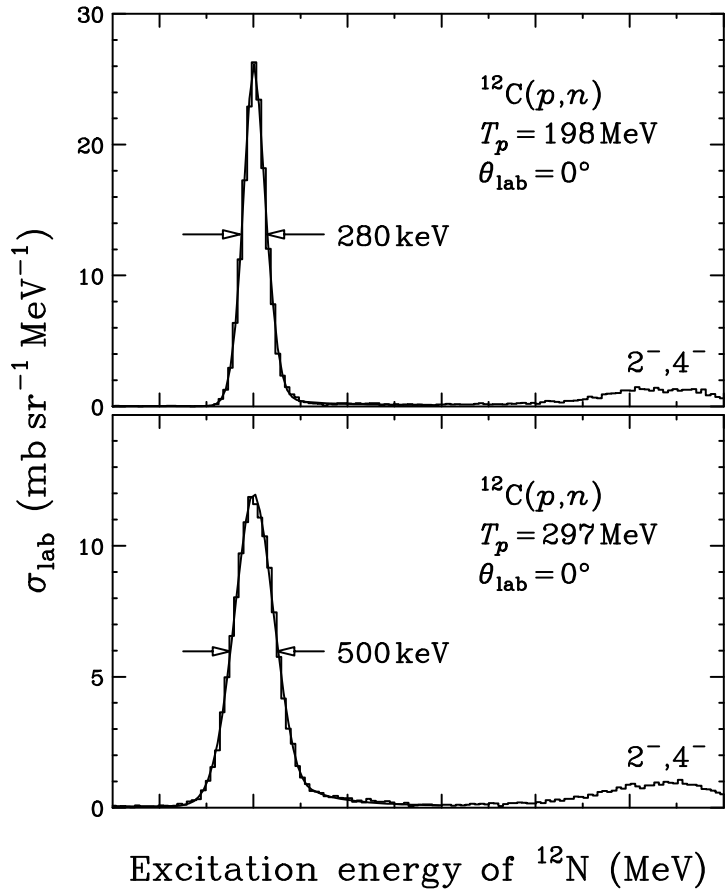}%
\end{center}
\caption{
 Measurements of the ${}^{12}{\rm C}(p,n)$ reaction 
at $T_p$ = 198 MeV (upper panel) and 297 (lower panel) MeV 
with 5 cm-thick plastic scintillation counters.
 The curve of each panel shows the reproduction of the 
${}^{12}{\rm N}(\mathrm{g.s.})$ state with the hyper-Gaussian peak.
\label{fig:expres}}
\end{figure}

 We have constructed the one-dimensional 
plastic scintillation counters with a size of 
100 cm $\times$ 10 cm  $\times$ 5cm.
 With these counters, we measured the 
${}^{12}{\rm C}(p,n)$ reaction at $\theta_{\mathrm{lab}}$ = $0^{\circ}$ 
and incident beam energies of $T_p$ = 198 and 295 MeV.
 The target thickness is 38 $\mathrm{mg/cm^{2}}$ whose 
contribution to the final energy resolution is negligible small.
 The results are shown in Fig.~\ref{fig:expres}.
 The energy resolutions were evaluated by fitting the 
peak of the ${}^{12}{\rm N}$ state to the standard hyper-Gaussian 
and they are about 280 and 500 keV FWHM for 
$T_p$ = 198 and 295 MeV data, respectively.
 These energy resolutions are consistent with the 
simulation results shown in Fig.~\ref{fig:deltatn}, and they 
satisfy our requirements.
 These counters have been used in the NPOL3 which 
will discussed in the next section.

\section{Neutron detector NPOL3 and its performance}
\label{sec:single}

\subsection{Neutron detector NPOL3}

\begin{figure}
\begin{center}
\includegraphics[width=0.9\linewidth,clip]{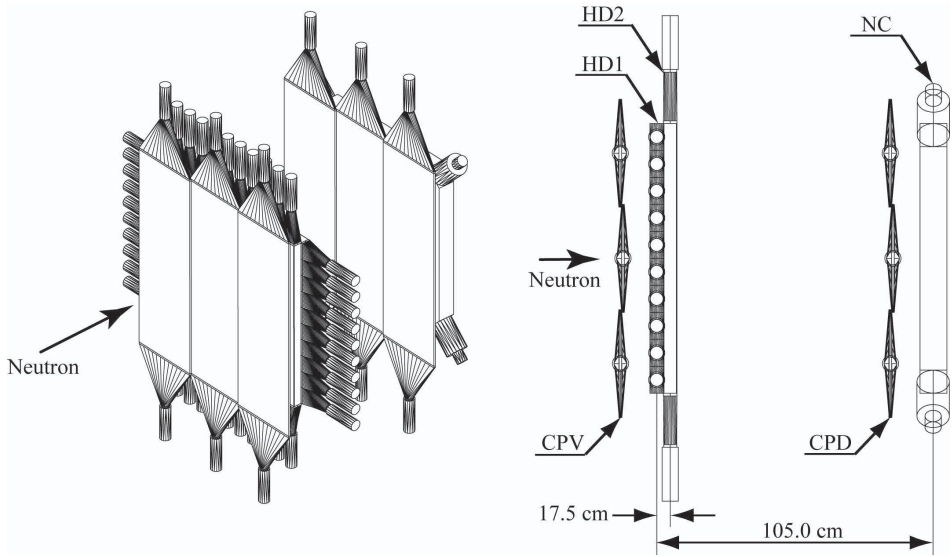}%
\end{center}
\caption{
 Schematic (left) and top (right) views of the NPOL3 system.
 In the polarimetry mode of NPOL3, 
HD1 and HD2 are the analyzer planes while NC is the catcher plane.
 Thin plastic scintillator planes are used to 
veto (CPV) or identify (CPD) charged particles.
\label{fig:npol3}}
\end{figure}

 The neutron detector NPOL3 has two main neutron 
detector planes, HD1 and HD2.
 Each plane consists of 10 stacked 100 cm $\times$ 10 cm  $\times$ 5cm
plastic scintillator bars made of Bicron BC408.
 Two thin plastic scintillator planes CPV and CPD, 
used to identify charged particles, complete the NPOL3 which 
is schematically shown in Fig.~\ref{fig:npol3}. 
 Additional two-dimensional neutron counter NC is used in polarimetry 
mode, which will be discussed in Sec.~\ref{sec:calib}.
 A scintillator in HD1 or HD2 is viewed by the fast Hamamatsu 
\cite{url_hamamatsu}
H2431 (HD1) or H1949 (HD2) PMTs attached at each end.
 The one-dimensional position is deduced from the fast timing 
information derived from PMT.
 The CPV and CPD are made of three sets of plastic scintillators 
(BC408) with a size of 102 cm $\times$ 35 cm $\times$ 0.5 cm.
 The scintillator is viewed from both sides (U:up and D:down) by 
the Hamamatsu H7195 PMT through fish-tail shape light guides.
 The CPV is used to veto charged particles entered into the NPOL3, 
and the CPD is used to detect the recoiled charged particles.
 
 In the neutron detector mode of NPOL3, a neutron is 
detected by either neutron detector plane HD1 or HD2.
 Furthermore we have required that the recoiled charged particle 
should be detected by the following detector plane.
 This means that we measure neutrons in the polarimetry (double-scattering)
mode of the NPOL3 system.
 This procedure has been applied in the data analysis of the 
INPOL system \cite{nima_457_309_2001}, and it is useful 
to improve the final energy resolution as well as to
reduce the low-energy tail component.
 Thus we have applied this procedure in our data analysis.

\subsection{Light output calibration}

 The neutron detection efficiency depends on the threshold
level $\mathcal{L}_{\mathrm{th}}$ of the scintillator light 
output.
 The total light output $\overline{\mathcal{L}}$ is constructed 
by taking the geometric mean of the PMT outputs at both ends, 
$\mathcal{L}_L$ and $\mathcal{L}_R$, as
\begin{equation}
\overline{\mathcal{L}} = \sqrt{\mathcal{L}_L\cdot\mathcal{L}_R}.
\end{equation}
 The light output $\overline{\mathcal{L}}$ is calibrated by using 4.4 MeV 
$\gamma$-rays from a ${}^{241}{\rm Am}$-${}^{9}{\rm Be}$ neutron 
source and cosmic rays (mostly $\mu^\pm$).
 In the cosmic-ray measurement, 10-bar hits in either HD1 or HD2 
were required.
 Thus horizontal and zenith angles can be measured in HD1 and HD2, 
respectively, on the assumption that cosmic rays pass 
in straight lines through the detector plane.
 The $\overline{\mathcal{L}}$ for cosmic rays has been corrected 
for these angles in order to evaluate the $\overline{\mathcal{L}}$
in case cosmic rays pass the scintillator of 10 cm in thickness.

\begin{figure}
\begin{center}
\includegraphics[width=0.7\linewidth,clip]{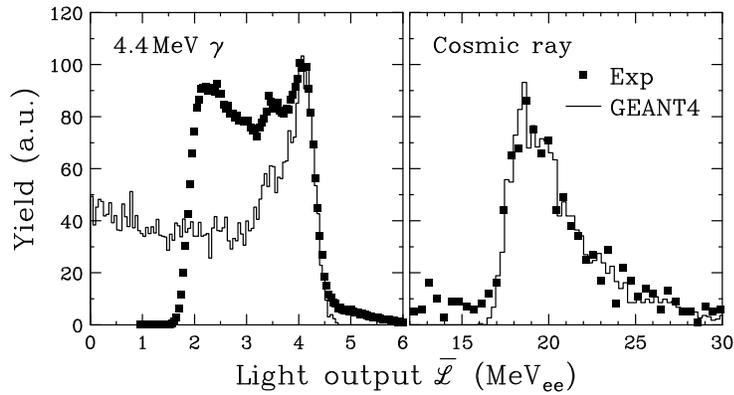}%
\end{center}
\caption{
 Light output spectra for 4.4 MeV $\gamma$-rays from 
a ${}^{241}{\rm Am}$-${}^{9}{\rm Be}$ neutron source (left panel) 
and cosmic rays (right panel).
 The histograms represent the results of the Monte-Carlo
simulations with {\sc geant4}.
 See text for details.
\label{fig:madc}}
\end{figure}

 The left and right panels of Fig.~\ref{fig:madc} show 
the $\overline{\mathcal{L}}$ spectra 
for 4.4 MeV $\gamma$-rays and cosmic rays, respectively.
 We can clearly see the Compton edge (4.2 $\mathrm{MeV_{ee}}$) 
for the $\gamma$-ray spectrum and the sharp peak with a Landau tail 
for the cosmic-ray spectrum.
 The histograms represent the results of the Monte-Carlo 
simulations with {\sc geant4} described in Sec.~\ref{sec:imp}.
 The simulations successfully reproduce the main components 
of the Compton edge and the sharp peak 
in $\gamma$-ray and cosmic-ray spectra, respectively.
 Furthermore they well describe the 
one-photon escape peak at $\sim$3.5 $\mathrm{MeV_{ee}}$ 
in the $\gamma$-ray spectrum 
and the Landau tail in the cosmic-ray spectrum.
 Note that the discrepancy between the data and the 
simulation result for 4.4 MeV $\gamma$-rays in the low 
$\overline{\mathcal{L}}$ region is due to the contributions from 
$\gamma$-rays other than 4.4 MeV $\gamma$-rays from the neutron source 
which are not taken into account in the simulations.

\subsection{Neutron detection efficiency}

 The laboratory differential cross section 
$d\sigma_{\rm lab}/d\Omega$ is related to the 
number of measured neutrons $N_n$ as
\begin{equation}
\frac{d\sigma_{\rm lab}}{d\Omega}=
\frac{N_n}{N_p\rho \Delta\Omega \epsilon T f_{\rm live}},
\end{equation}
where $N_p$ is the number of incident protons, 
$\rho$ is the target density,
$\Delta\Omega$ is the solid angle,
$\epsilon$ is the intrinsic neutron detection efficiency,
$T$ is the transmission factor along the neutron flight path, 
and $f_{\rm live}$ is the experimental live time.

 We measured the the product $\epsilon T$ by measuring the
neutron yield from the $0^{\circ}$ 
${}^{7}{\rm Li}(p,n){}^{7}{\rm Be}$(g.s. + 0.43 MeV) reaction
which has a constant center of mass (c.m.) cross section 
of $\sigma_{\rm c.m.}(0^{\circ})$ = 27.0 $\pm$ 0.8 mb/sr 
at the bombarding energy range of $T_p$ = 80--795 MeV 
\cite{prc_41_2548_1990}.
 The measurements were performed at $T_p$ = 198 and 295 MeV 
by using the enriched ${}^{7}{\rm Li}$ target with a thickness
of 54 $\mathrm{mg/cm^2}$.
 The integrated beam current $N_p$ was measured by using an 
electrically isolated graphite beam stop (Faraday cup) connected 
to a current integrator.
 The live time $f_{\rm live}$ was typically 90\%.

\begin{figure}
\begin{center}
\includegraphics[width=0.7\linewidth,clip]{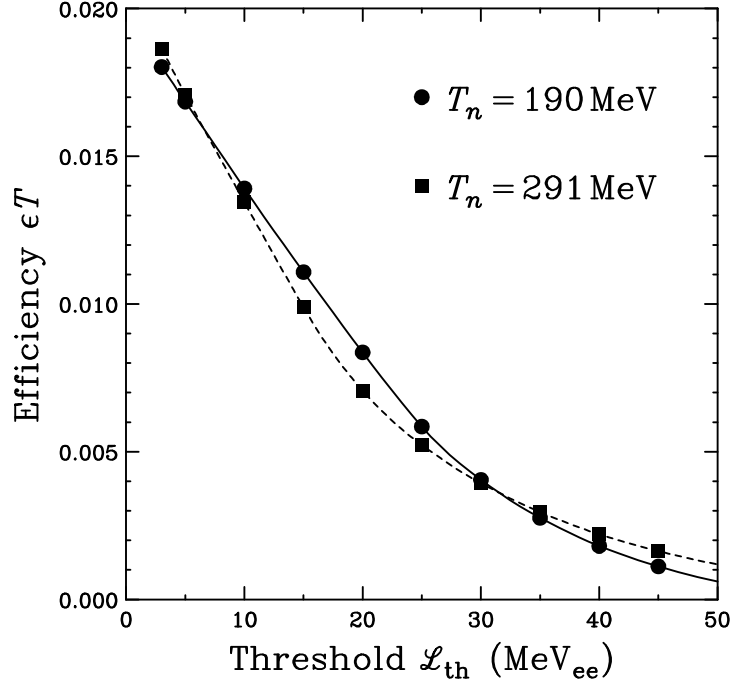}%
\end{center}
\caption{
 Neutron detection efficiencies of 5-cm thick plastic 
scintillation counters as a function of detection 
threshold $\overline{\mathcal{L}}_{\mathrm{th}}$.
 The filled circles and the filled boxes represent the results 
at $T_n$ = 190 and 291 MeV, respectively.
\label{fig:eff}}
\end{figure}

 The finally obtained neutron detection efficiencies 
($\epsilon T$) are shown in Fig.~\ref{fig:eff} as a function of the 
light output threshold $\overline{\mathcal{L}}_{\rm th}$.
 The systematic uncertainty is estimated to be about 6\% by 
taking into account of the uncertainties both of the 
${}^{7}{\rm Li}(p,n)$ cross section and of the 
${}^{7}{\rm Li}$ target thickness.
 It is found that the detection efficiencies are almost 
independent of the neutron kinetic energy with a value 
of $\sim$0.017 at $\overline{\mathcal{L}}_{\mathrm{th}}$ 
= 5 $\mathrm{MeV_{ee}}$.

\section{Neutron polarimeter NPOL3 and calibration procedure}
\label{sec:calib}

\subsection{Neutron polarimeter NPOL3}   

 In the polarimetry mode of NPOL3, the neutron 
detector planes, HD1 and HD2, serve as neutron 
polarization analyzers, and the following two-dimensional 
position-sensitive neutron counter NC acts as a catcher of 
doubly scattered neutrons or recoiled protons.
 Neutron polarization is determined from the asymmetry 
of the $\vec{n}+p$ events whose analyzing powers can be 
rigorously obtained by using the phase-shift analysis of the 
NN scattering.
 The $\vec{n}+p$ events are kinematically resolved from the 
other events by using time, position, and pulse-height 
information from both analyzer and catcher planes.
 Note that this kinematical selection significantly reduces 
the background events from the wraparound of slow neutrons from 
preceding beam pulses, cosmic rays, and the target gamma rays.
 However, the quasielastic $\vec{n}+\mathrm{C}$ reaction has an 
influence on the determination of the neutron polarization 
because its kinematical condition is similar to that of the 
$\vec{n}+p$ scattering.
 Thus we have measured the effective analyzing powers which 
include the contributions from both $\vec{n}+p$ and 
$\vec{n}+\mathrm{C}$ by using the polarized neutron beams.

\subsection{Effective analyzing powers}

 There are two channels, $(\vec{n},n)$ and $(\vec{n},p)$, 
in the polarimetry mode.
 Doubly scattered neutrons and recoiled protons are measured 
by the catcher plane NC in $(\vec{n},n)$ and $(\vec{n},p)$ 
channels, respectively.
 The effective analyzing powers,
$A_{y\mathrm{;eff}}^{nn}$ and $A_{y\mathrm{;eff}}^{np}$, 
of $(\vec{n},n)$ and 
$(\vec{n},p)$ channels can be determined by using the neutron beam 
with a known polarization $p_N'$.
 $A_{y\mathrm{;eff}}^{nn}$ and $A_{y\mathrm{;eff}}^{np}$ are 
deduced by
\begin{equation}
A_{y;\mathrm{eff}}^{nn} = \frac{\epsilon^{nn}}{p_N'}, \quad
A_{y;\mathrm{eff}}^{np} = \frac{\epsilon^{np}}{p_N'},
\end{equation}
where $\epsilon^{nn}$ and $\epsilon^{np}$ are the asymmetries 
measured for $(\vec{n},n)$ and $(\vec{n},p)$ channels, respectively.

\subsection{Polarized neutron beams}

 The first source of polarized neutrons is the 
$0^{\circ}$ ${}^{2}{\rm H}(\vec{p},\vec{n})pp$ reaction 
at $T_p$ = 198 MeV.
 The $D_{LL}$ value at $T_p$ = 198 MeV was obtained by 
using the data in Ref.~\cite{prl_71_684_1993}.
 Because this reaction is mainly a Gamow-Teller transition, 
the polarization transfer coefficients $D_{ij}$ satisfy 
\cite{jpsj_73_1611_2004} 
\begin{equation}
D_{NN}(0^{\circ})=-\frac{1+D_{LL}(0^{\circ})}{2}.
\label{eq:dij_gt}
\end{equation}
 The neutron polarization $p_N'$ is given by using 
$D_{NN}(0^{\circ})$ as 
\begin{equation}
p_N' = p_N D_{NN}(0^{\circ}),
\end{equation}
where $p_N$ is the polarization of the incident proton 
beam.

 The second source is the $0^{\circ}$ 
${}^{6}{\rm Li}(\vec{p},\vec{n}){}^{6}{\rm Be}({\rm g.s.})$ 
reaction at $T_p$ = 198 MeV.
 This reaction is also a GT transition, and its $D_{NN}$ 
value was measured at $T_p$ = 200 MeV by Taddeucci 
\cite{cnj_65_557_1987}.
 The result can be used as a double-check of the calibration 
performed by using the neutrons from ${}^{2}{\rm H}(p,n)pp$.

 The third source is the $0^{\circ}$ 
${}^{12}{\rm C}(\vec{p},\vec{n}){}^{12}{\rm N}({\rm g.s.})$ 
reaction at $T_p$ = 198 MeV.
 The $D_{NN}(0^{\circ})$ value of this GT transition was also measured 
at $T_p$ = 200 MeV by Taddeucci \cite{cnj_65_557_1987}.
 Unfortunately, its uncertainty is relatively large compared 
with that of the ${}^{2}{\rm H}(\vec{p},\vec{n})$ or
${}^{6}{\rm Li}(\vec{p},\vec{n}){}^{6}{\rm Be}({\rm g.s.})$ reaction.
 Thus the final accuracy of $A_{y;\mathrm{eff}}$ is limited 
by the uncertainty of the neutron beam polarization originating 
from the uncertainty of $D_{NN}(0^{\circ})$,
 Therefore we have used the following method which does not 
require to know the $D_{ii}(0^{\circ})$ values.

 The GT transition is the spin-flip and unnatural-parity 
transition, and its polarization transfer coefficients 
satisfy the relation in Eq.~(\ref{eq:dij_gt}).
 Here we assume that two kinds of polarized proton beams are 
available; one has pure longitudinal polarization $p_L$ and 
the other has pure normal polarization $p_N$.
 The neutron polarizations at $0^{\circ}$ become
$p_L'$ = $p_L D_{LL}(0^{\circ})$ and $p_N'$ = $p_N D_{NN}(0^{\circ})$ 
for the beam polarizations of $p_L$ and $p_N$, respectively.
 Then the resulting asymmetries measured by a neutron polarimeter are 
\begin{equation}
\begin{split}
\epsilon_L &= p_L'A_{y;\mathrm{eff}} 
  = p_LD_{LL}(0^{\circ})A_{y;\mathrm{eff}},\\
\epsilon_N &= p_N'A_{y;\mathrm{eff}} 
  = p_ND_{NN}(0^{\circ})A_{y;\mathrm{eff}}.
\end{split}
\label{eq:asy_vs_ayeff}
\end{equation}
 By using Eqs.~(\ref{eq:dij_gt}) and (\ref{eq:asy_vs_ayeff}), 
$A_{y;\mathrm{eff}}$ can be expressed as
\begin{equation}
A_{y;\mathrm{eff}} = -\left(
  \frac{\epsilon_L}{p_L}
+2\frac{\epsilon_N}{p_N}
\right).
\end{equation}
 Thus the $A_{y;\mathrm{eff}}$ value can be obtained without 
knowing a priori the values of $D_{ii}(0^{\circ})$, and we have 
applied this technique to the 
${}^{12}{\rm C}(\vec{p},\vec{n}){}^{12}{\rm N}({\rm g.s.})$ 
reaction.

\subsection{Experimental conditions}

 We used a deuterated polyethylene ($\mathrm{CD_2}$) target 
with a thickness of 228 $\mathrm{mg/cm^2}$ as a deuteron target.
 Neutrons from the ${}^{2}{\rm H}(\vec{p},\vec{n})pp$ reaction
were clearly separated from those from the 
${}^{12}{\rm C}(\vec{p},\vec{n}X)$ reaction 
since the former reaction $Q$ value of $-2.2$ MeV is significantly 
smaller than the latter $Q$ value  of $\ge$ $-18.1$ MeV.
 We also used 99\% enriched ${}^{6}{\rm Li}$ and natural C targets 
with thicknesses of 181 and 38 $\mathrm{mg/cm^2}$, respectively.
 The proton beam intensity and its polarization are 
typically 80 nA and 0.70, respectively.

\subsection{Sector methods}

\begin{figure}
\begin{center}
\includegraphics[width=0.7\linewidth,clip]{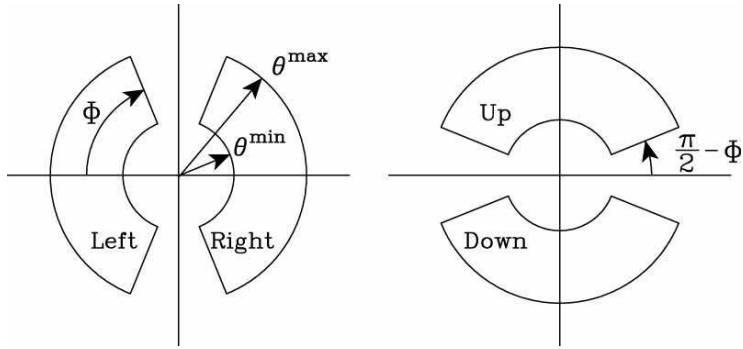}%
\end{center}
\caption{
 Illustration of the sector definition on the 
catcher plane NC for left/right (left panel) 
and up/down (right panel) asymmetries. 
\label{fig:sector}}
\end{figure}

 We have adopted the sector method in order to obtain 
the asymmetry.
 In this method, the plane $(\theta,\phi)$ defined 
by the scattering polar angle 
$\theta$ and the azimuthal angle $\phi$ is divided into four sectors:
Left, Right, Up, and Down, as is shown in Fig.~\ref{fig:sector}.
 Each sector is defined by scattering angle limits
 $\theta^{\mathrm{min}}$ and $\theta^{\mathrm{max}}$ and 
the azimuthal half-angle $\Phi$.

 The asymmetry $\epsilon$ is calculated by using the 
event numbers in these sectors.
 For example, the relation between the left-right asymmetry 
$\epsilon_N'$ and the event numbers is as follows.
 The azimuthal distribution of scattered particles can be 
described as 
\begin{equation}
f(\phi) = \frac{1}{2\pi}(1+p_N'\cos\phi),
\end{equation}
where we suppress the $\theta$ dependence for simplicity.
 Then the event numbers scattered to left and right become
\begin{equation}
\begin{split}
Y_L &= \int_{-\Phi}^{\Phi}Nf(\phi)d\phi = 
\frac{N}{\pi}(\Phi + \epsilon_{N'}\sin\Phi),\\
Y_R &= \int_{\pi-\Phi}^{\pi+\Phi}Nf(\phi)d\phi = 
\frac{N}{\pi}(\Phi - \epsilon_{N'}\sin\Phi),
\end{split}
\end{equation}
where $N$ is the event number accepted in the $\theta$ bin, 
and the $\phi$ acceptance is assumed to be constant.
 Thus the asymmetry $\epsilon_N'$ = $p_N'A_{y;\mathrm{eff}}$ 
is found to be 
\begin{equation}
\epsilon_N' = \frac{Y_L-Y_R}{Y_L+Y_R}\frac{\Phi}{\sin\Phi}.
\label{eq:asy_def}
\end{equation}
 Note that the sideway polarization $p_S'$ is deduced from the 
up-down asymmetry $\epsilon_S'$ which is obtained by substituting 
$L\rightarrow D$ and $R\rightarrow U$ in Eq.~(\ref{eq:asy_def}).

\section{Optimization of performance of NPOL3}
\label{sec:opt}

\subsection{Optimization criteria}
\label{ssec:cri}

 The performance of a polarimeter can be measured by its 
figure-of-merit (FOM) value defined as
\begin{equation}
\mathrm{FOM} = N(A_{y;\mathrm{eff}})^2,
\end{equation}
where $N$ is the total event number $Y_L+Y_R$ or $Y_U+Y_D$.
 We can minimize the uncertainty of the final result
(neutron polarization and $D_{ij}$) by maximizing this 
FOM value.

 We have applied three types (Types-I, II, and III) 
of software cuts to the measured $\theta$, $\phi$, 
inter-plane velocity, and pulse height information 
of the NPOL3.
 The FOM value is maximized in Type-I, 
whereas the $A_{y;\mathrm{eff}}$ is maximized 
in Type-II with keeping the FOM value larger than 40\% of the
optimum value.
 In Type-III, both FOM and $A_{y;\mathrm{eff}}$ values 
have been optimized with keeping the FOM value larger than 
80\% of the optimum value.
 The Type-I is the best choice from the statistical point of 
view in order to maximize the performance of the NPOL3.
 However, in general, a higher $A_{y;\mathrm{eff}}$ value 
in Type-II or III is helpful to reduce the systematic uncertainty
of the final result.
 Thus we have obtained these three sets and we will select 
the optimum set depending on the experimental condition.

 In the following, we show the results by using the 
data of the ${}^{2}{\rm H}(\vec{p},\vec{n})pp$ reaction 
because the total event number is significantly larger 
than that of other two polarized-neutron production reactions.
 The $A_{y;\mathrm{eff}}$ value is also obtained from the 
${}^{6}{\rm Li}(\vec{p},\vec{n}){}^{6}{\rm Be}(g.s.)$ data 
by applying the same software cuts.
 The mean neutron energy is about 193 MeV for both the 
the ${}^{2}{\rm H}(\vec{p},\vec{n})pp$ and 
${}^{6}{\rm Li}(\vec{p},\vec{n}){}^{6}{\rm Be}(g.s.)$ reactions.
 Thus we can use the results for a double check of the calibration.
 The same software cuts are also applied to the 
${}^{12}{\rm C}(\vec{p},\vec{n}){}^{12}{\rm N}(g.s.)$ data.
 The mean neutron energy of $\sim$180 MeV is significantly low, 
therefore, 
the result can be used to discuss the neutron energy dependence 
of $A_{y;\mathrm{eff}}$.

\subsection{Optimization of $\Phi$}

\begin{figure}
\begin{center}
\includegraphics[width=0.9\linewidth,clip]{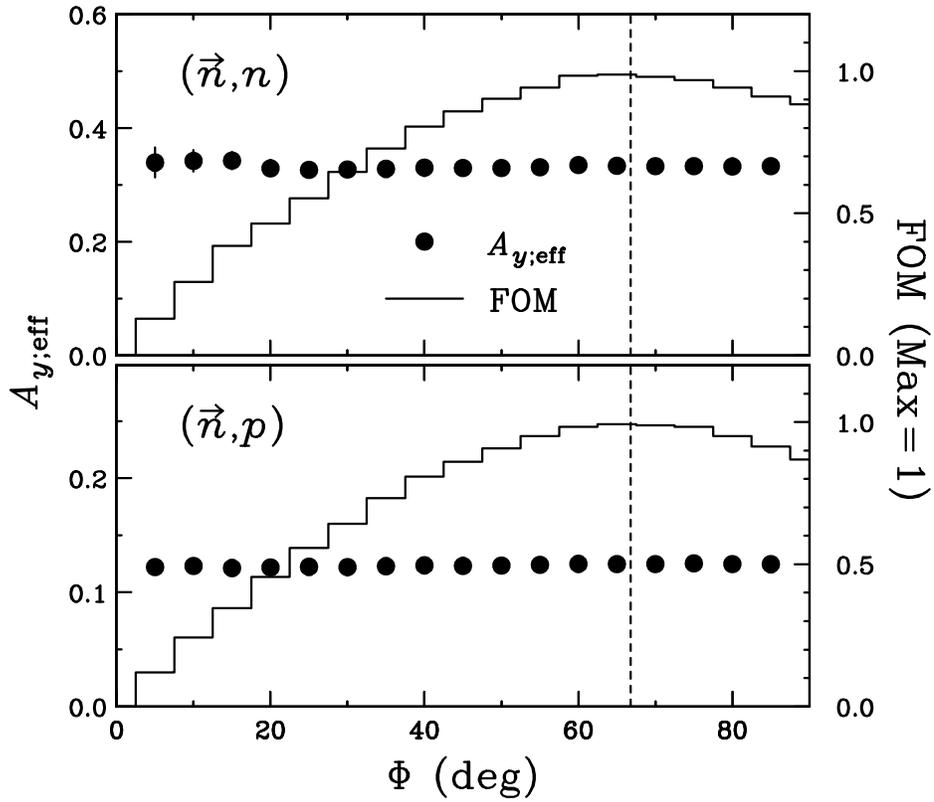}%
\end{center}
\caption{
 Effective analyzing powers $A_{y;\mathrm{eff}}$ and FOM 
as a function of azimuthal half-angle $\Phi$.
 The upper and lower panels correspond to the results 
for $(\vec{n},n)$ and $(\vec{n},p)$ channels, respectively.
\label{fig:phi}}
\end{figure}

 Figure~\ref{fig:phi} shows the $A_{y;\mathrm{eff}}$ and FOM values 
in Type-I as a function of the azimuthal half-angle $\Phi$.
 The FOM values are normalized to a maximum FOM value of 1.
 The $A_{y;\mathrm{eff}}$ values are 
insensitive to the choice of $\Phi$, whereas the FOM values 
take a maximum at $\Phi$ $\sim$ $67^{\circ}$.
 This angle is consistent with the simple estimation of the 
optimum $\Phi$ = $66.8^{\circ}$ \cite{nima_404_355_1998}.
 Because the $\Phi$ dependence is almost independent of
the software cut types, we set $\Phi$ = $66.8^{\circ}$ 
in all cases.

\subsection{Optimization of $\theta$ window} 

\begin{figure}
\begin{center}
\includegraphics[width=0.9\linewidth,clip]{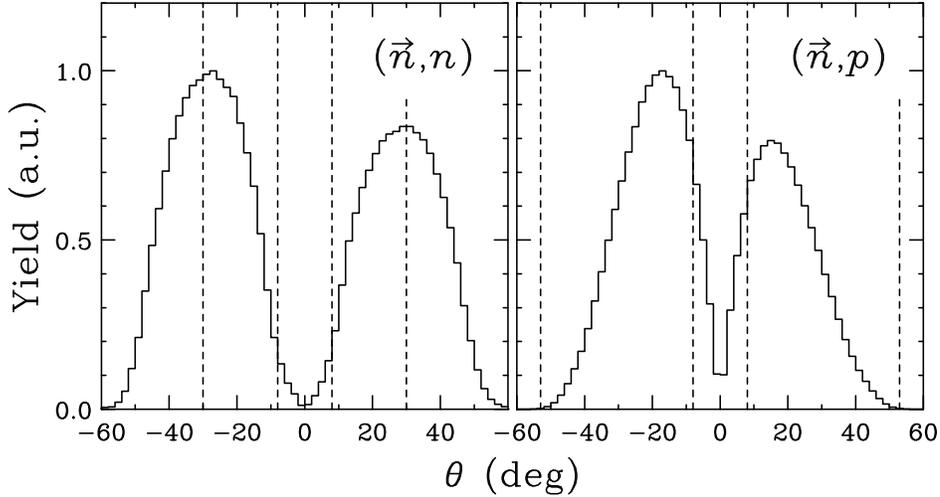}%
\end{center}
\caption{
 Distributions of the doubly scattered particles as 
a function of scattering polar angle $\theta$.
 The left and right panels correspond to the results 
for $(\vec{n},n)$ and $(\vec{n},p)$ channels, respectively.
\label{fig:theta1d}}
\end{figure}

 The event selection on $\theta$ is necessary to prevent 
dilution of $A_{y;\mathrm{eff}}$ by events with small or negative 
analyzing powers.
 Figure~\ref{fig:theta1d} shows the scattered-particle distribution 
as a function of $\theta$ in Type-I.
 The results in other Types-II and III are also identical.
 The double scattering events with a scattering angle 
up to $\sim$ $60^{\circ}$ are accepted by the NPOL3.
 The analyzing power $A_y(\theta)$ of the free 
$\vec{n}+p$ scattering is positive only up to $45^{\circ}$ 
and $41^{\circ}$ for $(\vec{n},n)$ and $(\vec{n},p)$ channels, 
respectively, at $T_n$ = 193 MeV.
 Thus part of the events should be rejected on the basis 
of small $A_y$ as described below.

\begin{figure}
\begin{center}
\includegraphics[width=0.9\linewidth,clip]{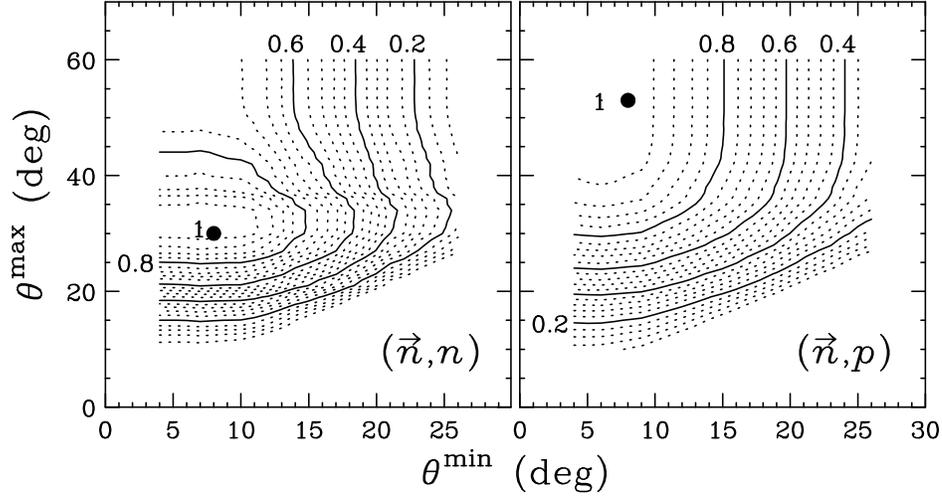}%
\end{center}
\caption{
 FOM of NPOL3 as functions of lower and upper limits of the 
scattering polar angle window 
$(\theta^{\mathrm{min}},\theta^{\mathrm{max}})$.
 The left and right panels correspond to the results 
for $(\vec{n},n)$ and $(\vec{n},p)$ channels, respectively.
 The selected limits in Type-I are indicated by the 
filled circles.
\label{fig:theta2d}}
\end{figure}

\begin{figure}
\begin{center}
\includegraphics[width=0.9\linewidth,clip]{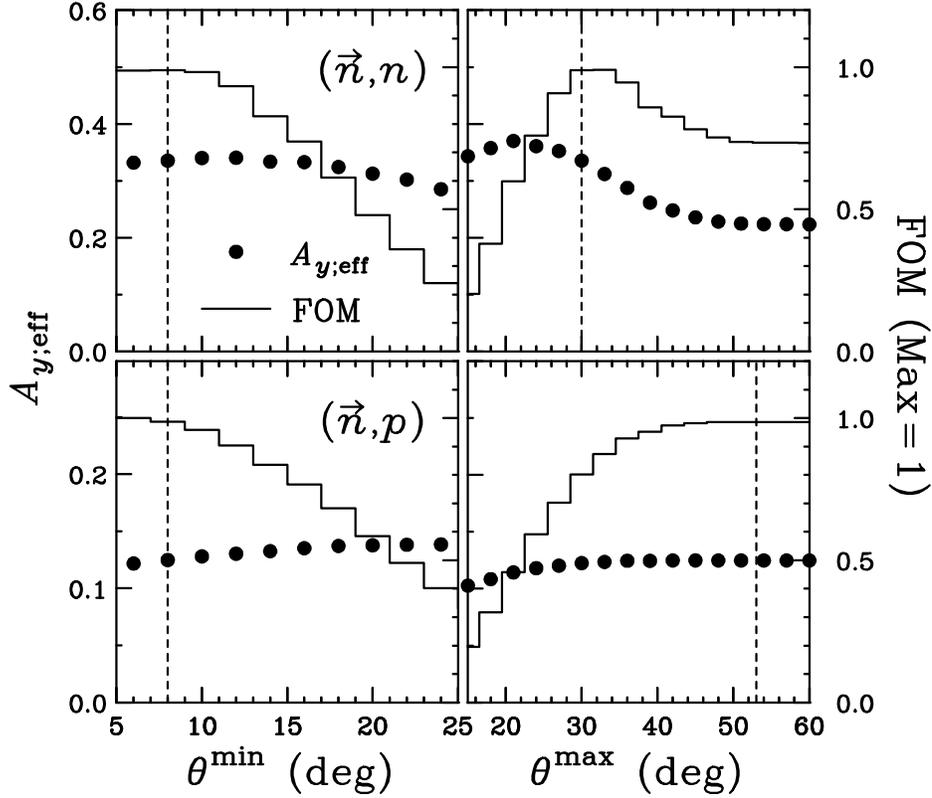}%
\end{center}
\caption{
 Effective analyzing powers $A_{y;\mathrm{eff}}$ and FOM 
as a function of either lower (left panels) or upper (right panels)
limit of the scattering polar angle $\theta$.
 The upper and lower panels correspond to the results 
for $(\vec{n},n)$ and $(\vec{n},p)$ channels, respectively.
\label{fig:theta}}
\end{figure}

 Figure~\ref{fig:theta2d} shows the contour plots of the FOM values 
in Type-I as functions of both $\theta^{\mathrm{min}}$ and 
$\theta^{\mathrm{max}}$.
 Figure~\ref{fig:theta} represents the $A_{y;\mathrm{eff}}$ and FOM 
values as a function of either $\theta^{\mathrm{min}}$ or
$\theta^{\mathrm{max}}$.
 In both figures, the FOM values are normalized to a 
maximum FOM value of 1.
 The FOM values strongly depend on both $\theta^{\mathrm{min}}$
and $\theta^{\mathrm{max}}$, whereas the $A_{y;\mathrm{eff}}$ 
values weakly depend on them except for $\theta^{\mathrm{max}}$
in the $(\vec{n},n)$ channel.
 In cases of Types-II and III, the optimal 
$(\theta^{\mathrm{min}},\theta^{\mathrm{max}})$ values are 
selected on basis of the criteria described in Sec.~\ref{ssec:cri}.
 In Table~\ref{table:cut}, we present the 
finally chosen $(\theta^{\mathrm{min}},\theta^{\mathrm{max}})$
values.

\subsection{Optimization of $R_v$ window} 

\begin{figure}
\begin{center}
\includegraphics[width=0.9\linewidth,clip]{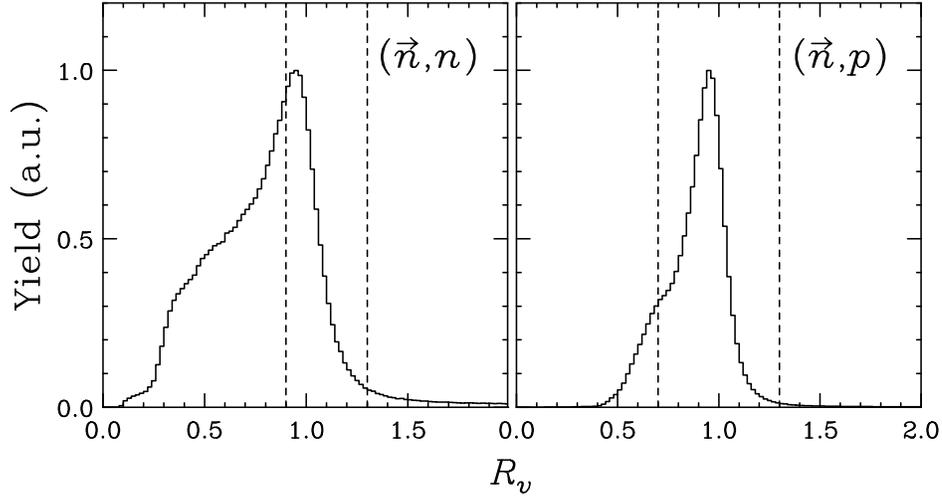}%
\end{center}
\caption{
 Distributions of the doubly scattered particles as 
a function of velocity ratio $R_v$.
 The left and right panels correspond to the results 
for $(\vec{n},n)$ and $(\vec{n},p)$ channels, respectively.
\label{fig:rv1d}}
\end{figure}

 The kinematical selection of the $\vec{n}+p$ events 
is performed by using the kinematical quantity 
$R_v$ defined as 
\begin{equation}
R_v = \frac{v}{v_{NN}},
\end{equation}
where $v$ is the measured particle velocity between the 
analyzer and catcher planes and $v_{NN}$ is the predicted 
velocity based on the NN kinematics.
 The $v_{NN}$ value is given by
\begin{equation}
v_{NN} = c \frac{2\cos\theta\sqrt{T_n/(T_n+2m_N)}
}{1-\cos^2\theta\,T_n/(T_n+m_N)},
\end{equation}
where $c$ is the light velocity and $m_N$ is the nucleon mass.
 Figure~\ref{fig:rv1d} shows the scattered-particle distribution 
in Type-I as a function of $R_v$.
 The peaks at $R_v$ $\sim$ 1 correspond to the events 
from the $p(\vec{n},n)p$ or $p(\vec{n},p)n$ reaction.
 The shoulders at $R_v < 1$ are due to the contributions 
from quasielastic scattering, inelastic scattering, and 
wraparounds.

\begin{figure}
\begin{center}
\includegraphics[width=0.9\linewidth,clip]{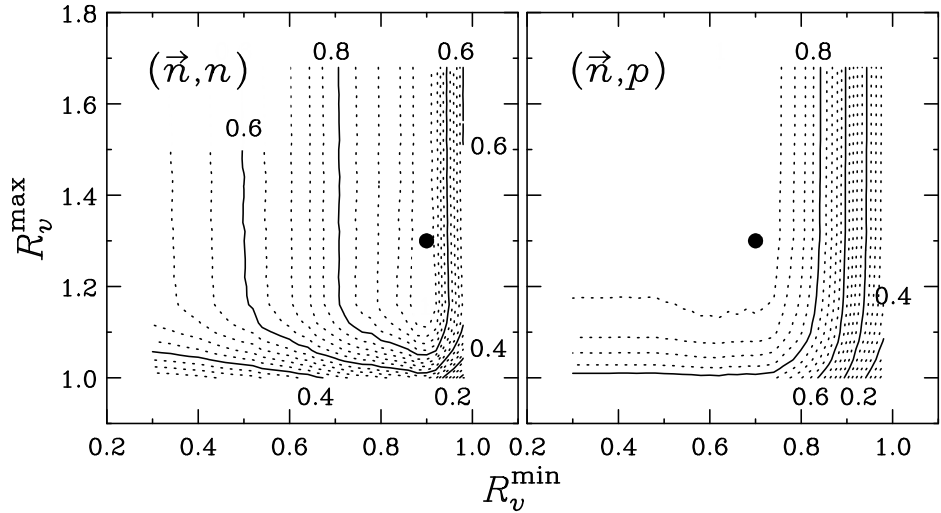}%
\end{center}
\caption{
 FOM of NPOL3 as functions of lower and upper limits of the 
velocity ratio window 
$(R_v^{\mathrm{min}},R_v^{\mathrm{max}})$.
 The left and right panels correspond to the results 
for $(\vec{n},n)$ and $(\vec{n},p)$ channels, respectively.
 The selected limits in Type-I are indicated by the 
filled circles.
\label{fig:rv2d}}
\end{figure}

\begin{figure}
\begin{center}
\includegraphics[width=0.9\linewidth,clip]{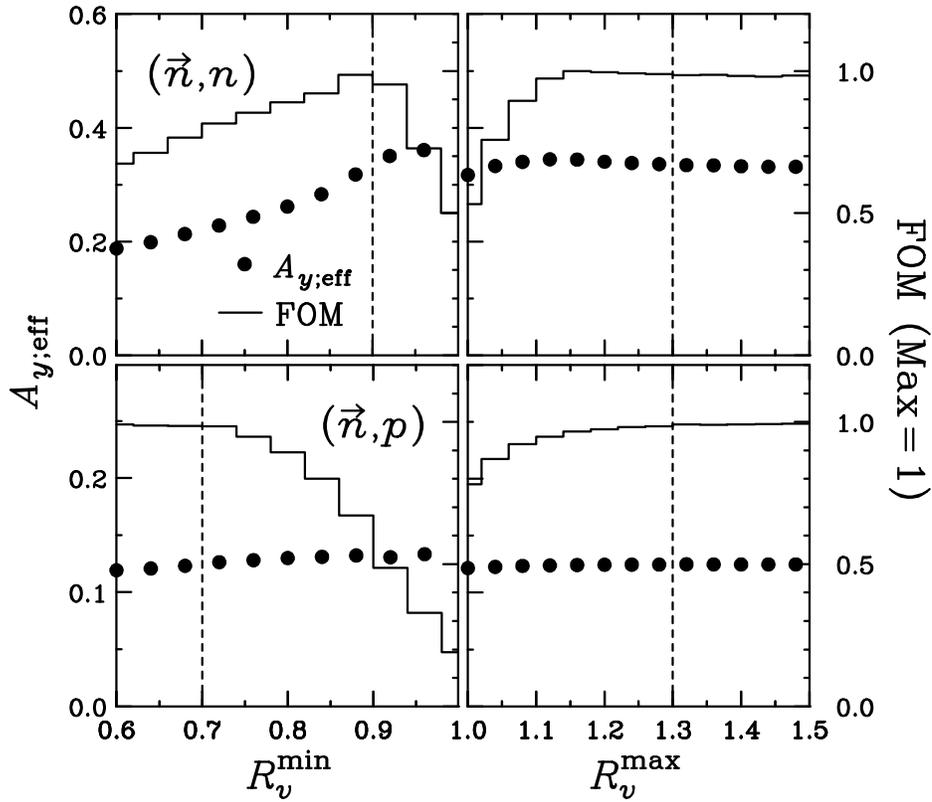}%
\end{center}
\caption{
 Effective analyzing powers $A_{y;\mathrm{eff}}$ and FOM 
as a function of either lower (left panels) or upper (right panels)
limit of the velocity ratio $R_v$.
 The upper and lower panels correspond to the results 
for $(\vec{n},n)$ and $(\vec{n},p)$ channels, respectively.
\label{fig:rv}}
\end{figure}

 Figure~\ref{fig:rv2d} shows the contour plots of the FOM values 
as functions of both $R_v^{\mathrm{min}}$ and $R_v^{\mathrm{max}}$.
 Figure~\ref{fig:rv} represents the $A_{y;\mathrm{eff}}$ and FOM 
values as a function of either 
$R_v^{\mathrm{min}}$ or $R_v^{\mathrm{max}}$.
 The FOM values are normalized to a maximum FOM value of 1 
in both figures.
 It is found that the FOM values are almost insensitive to 
$R_v^{\mathrm{max}}$ for $R_v^{\mathrm{max}}$ $>$ 1.3.
 However, $R_v^{\mathrm{max}}$ has been limited to be less 
than 1.3 to eliminated the contribution from $\gamma$-ray events 
with $R_v$ $>$ 1.5.
 In Types-II and III, the optimal 
$(R_v^{\mathrm{min}},R_v^{\mathrm{max}})$ values are 
selected on basis of the criteria described in Sec.~\ref{ssec:cri}.
 The final $(R_v^{\mathrm{min}},R_v^{\mathrm{max}})$
values chosen are listed in Table~\ref{table:cut}.

\subsection{Optimization of $\overline{\mathcal{L}}_{\mathrm{th}}$} 

\begin{figure}
\begin{center}
\includegraphics[width=0.9\linewidth,clip]{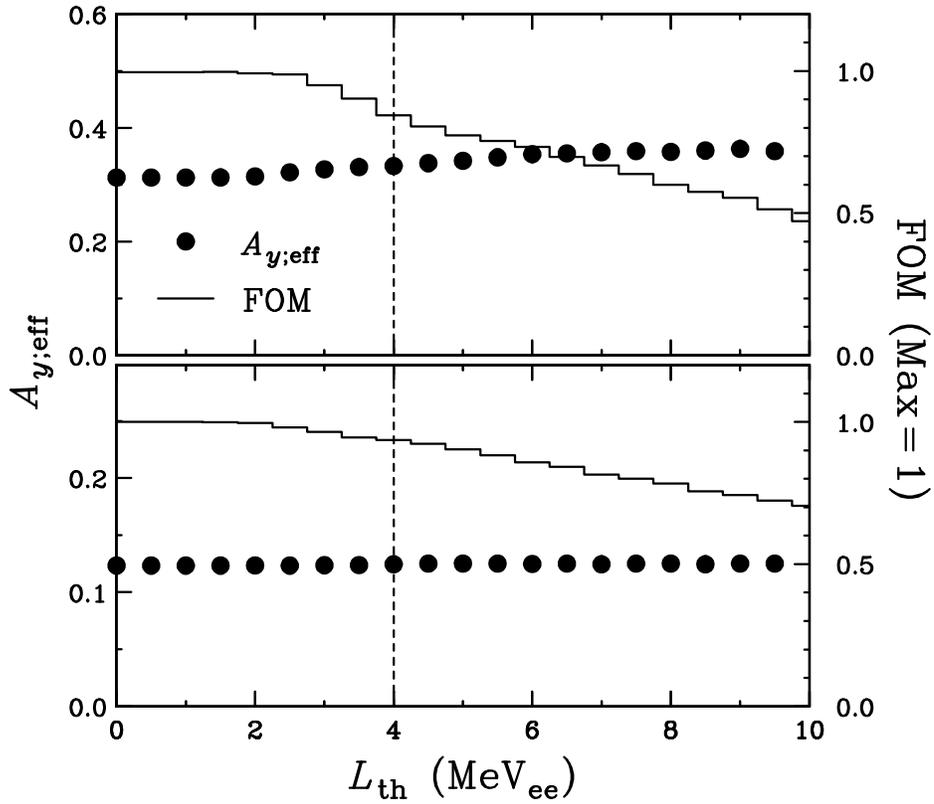}%
\end{center}
\caption{
 Effective analyzing powers $A_{y;\mathrm{eff}}$ and FOM 
as a function of detection threshold $\overline{\mathcal{L}}_{\mathrm{th}}$.
 The upper and lower panels correspond to the results 
for $(\vec{n},n)$ and $(\vec{n},p)$ channels, respectively.
\label{fig:lth}}
\end{figure}

 Figure~\ref{fig:lth} shows the $A_{y;\mathrm{eff}}$ and FOM 
values in Type-I as a function of $\overline{\mathcal{L}}_{\mathrm{th}}$.
 The FOM values monotonously decrease as increasing 
$\overline{\mathcal{L}}_{\mathrm{th}}$, whereas the $A_{y;\mathrm{eff}}$ 
values are almost independent of $\overline{\mathcal{L}}_{\mathrm{th}}$.
 In all software cut types, we have chosen
$\overline{\mathcal{L}}_{\mathrm{th}}$ = 4 $\mathrm{MeV}_{ee}$ to ensure 
reproducibility in the offline analysis rather than 
just
accept the hardware threshold of $\sim$2 $\mathrm{MeV}_{ee}$.

\section{Performance of polarimeter NPOL3} 
\label{sec:double}

\subsection{Effective analyzing powers}

 The effective analyzing powers $A_{y;\mathrm{eff}}$
obtained from the ${}^{2}{\rm H}(\vec{p},\vec{n})pp$  data
by applying previously described software cuts are summarized in 
Table~\ref{table:ayeff}.
 The $A_{y;\mathrm{eff}}$ values deduced from the  
${}^{6}{\rm Li}(\vec{p},\vec{n}){}^{6}{\rm Be}(\mathrm{g.s.})$  data
are also listed in Table~\ref{table:ayeff}.
 The results of the ${}^{2}{\rm H}(\vec{p},\vec{n})pp$ and 
${}^{6}{\rm Li}(\vec{p},\vec{n}){}^{6}{\rm Be}(\mathrm{g.s.})$ data 
are consistent with each other within their statistical uncertainties, 
which suggests the reliability 
of the present calibration procedure.

\begin{figure}
\begin{center}
\includegraphics[width=0.7\linewidth,clip]{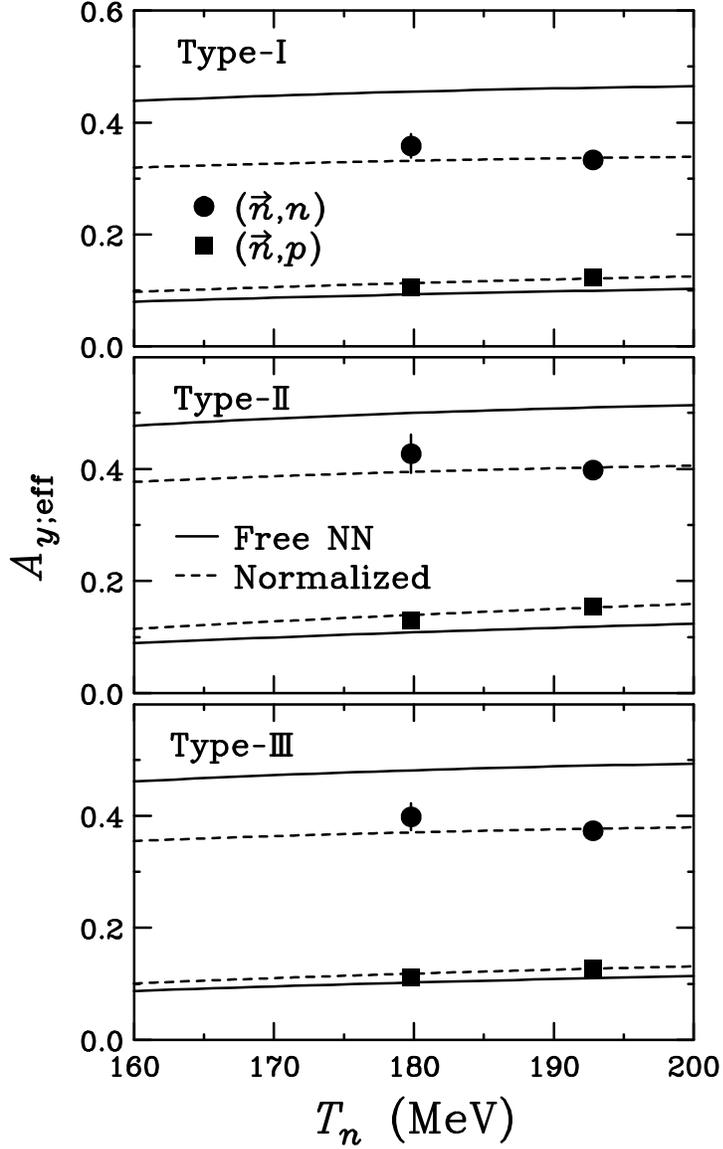}%
\end{center}
\caption{
 Effective analyzing powers of the NPOL3 system 
for $(\vec{n},n)$ (filled circles) and $(\vec{n},p)$ (filled boxes) 
channels, respectively, as a function of neutron energy.
 The top, middle, and bottom panels correspond to the results 
in the software cut Types-I, II, and III, respectively.
 The solid curves represent the calculations based on the 
free $\vec{n}+p$ observables.
 The dashed curves represent the calculations normalized to 
reproduce the experimental data.
\label{fig:ayeff}}
\end{figure}

 The $A_{y;\mathrm{eff}}$ values deduced from the  
${}^{12}{\rm C}(\vec{p},\vec{n}){}^{12}{\rm N}(\mathrm{g.s.})$  data
are plotted in Fig.~\ref{fig:ayeff}, and they are also 
listed in Table~\ref{table:ayeff}.
 The results of the ${}^{2}{\rm H}(\vec{p},\vec{n})pp$ data
are also displayed in Fig.~\ref{fig:ayeff}.
 The $A_{y;\mathrm{eff}}$ values have been estimated as a 
function of $T_n$ on the basis of the free $\vec{n}+p$ analyzing 
powers derived from the NN phase shift analysis \cite{said}.
 In the calculations, the $(\theta,\phi)$ cuts and the 
counter geometries have been properly
taken into account.
 The solid curves in Fig.~\ref{fig:ayeff} show the results of the 
calculations.
 The measured $A_{y;\mathrm{eff}}$ values are different from the 
calculated values by factors of 0.73--0.79 and 1.16--1.29 for 
$(\vec{n},n)$ and $(\vec{n},p)$ channels, respectively, 
depending on the event cut types.
 The difference of $A_{y;\mathrm{eff}}$ is mainly due to the
contributions from the quasielastic $\vec{n}+\mathrm{C}$ events.
 The dashed curves represent the calculated $A_{y;\mathrm{eff}}$ 
values normalized to reproduced the experimental results.
 These normalized values will be used 
in future data analysis 
in case there is no appropriate calibration result.

\subsection{Double scattering efficiency}

 The double scattering efficiencies $(\epsilon_{\mathrm{D.S.}})$ 
of the NPOL3 were deduced from the data of the 
${}^{2}{\rm H}(\vec{p},\vec{n})pp$ and 
${}^{12}{\rm C}(\vec{p},\vec{n}){}^{12}{\rm N}(\mathrm{g.s.})$ reactions.
 The results are listed in Table~\ref{table:eds}.
 The $\epsilon_{\mathrm{D.S.}}$ values in both $(\vec{n},n)$ 
and $(\vec{n},p)$ channels are almost constant in the present 
neutron energy region.

\section{Discussion}
\label{sec:discuss}

\subsection{$\theta$ dependence of $A_{y;\mathrm{eff}}$}

\begin{figure}
\begin{center}
\includegraphics[width=0.7\linewidth,clip]{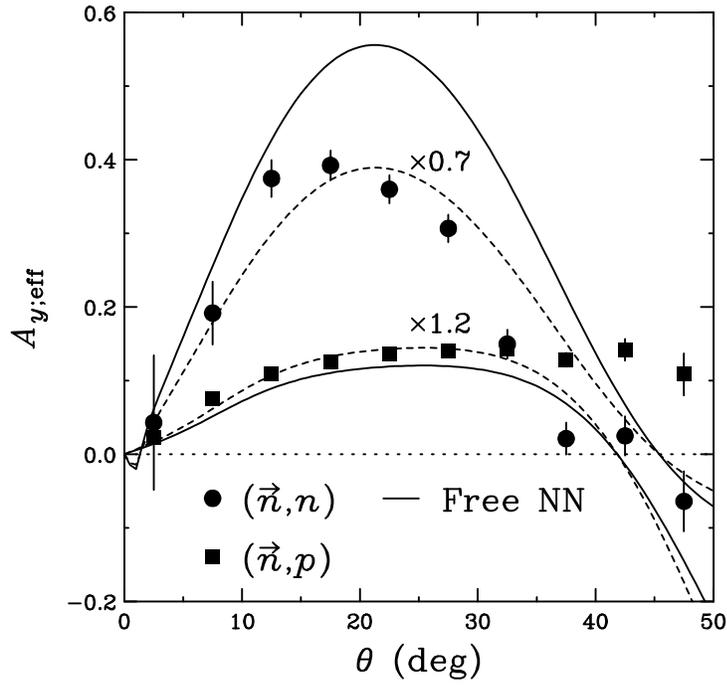}%
\end{center}
\caption{
 Angular distributions of effective analyzing powers of the NPOL3 system 
for $(\vec{n},n)$ (filled circles) and $(\vec{n},p)$ (filled boxes) 
channels, respectively.
 The solid curves represent the calculations based on the 
free $\vec{n}+p$ observables whereas the dashed curve represent the 
normalized results to reproduce the experimental data.
\label{fig:aytheta}}
\end{figure}

 In the sector method, the events within 
$(\theta^{\mathrm{min}},\theta^{\mathrm{max}})$ are integrated in 
order to evaluate the $A_{y;\mathrm{eff}}$ values in the calibration.
 Thus the results can be considered as the mean effective 
analyzing powers in the range of 
$\theta$ = $\theta^{\mathrm{min}}$--$\theta^{\mathrm{max}}$.
 We deduced the $A_{y;\mathrm{eff}}$ values as a function of 
$\theta$ in order to check whether the $\vec{n}+p$ events 
are properly selected in our analysis.
 As a typical example, 
the results in Type-III are shown in Fig.~\ref{fig:aytheta}.
 The solid curves display the angular distributions of 
$A_y(\theta)$ of the free $\vec{n}+p$ scattering.
 The measured angular distributions are very similar to those 
of the free $\vec{n}+p$ scattering.
 However the magnitudes are different mainly due to the 
contributions from the quasielastic $\vec{n}+\mathrm{C}$ events.
 The dashed curves are the results normalized to reproduce the 
measured values, and the normalization factors are 0.7 and 1.2 
for $(\vec{n},n)$ and $(\vec{n},p)$ channels, respectively.
 The agreement of the angular distributions supports the 
proper kinematical selection of the $\vec{n}+p$ events 
in our analysis.

\subsection{Comparison with INPOL system}

 Here we compare the performance of the NPOL3 system with that 
of the INPOL system \cite{nima_457_309_2001} which has been used 
in the same energy region.
 For the $(\vec{n},n)$ channel, the NPOL3 has $\sim$$1/2$ volume 
for both analyzer and catcher planes compared with the INPOL.
 Thus the double scattering efficiency $\epsilon_{\mathrm{D.S}}$ and 
the corresponding FOM value of the NPOL3
is expected to be $\sim$$1/4$ of those of the INPOL.
 The FOM value of the NPOL3 is $0.28\times 10^{-4}$ in Type-I 
at $T_n$ = 193 MeV, which is about $1/4$ of that of the INPOL 
reported as $0.92\times 10^{-4}$ at $T_n$ = 194 MeV.
 For the $(\vec{n},p)$ channel, the FOM value of the NPOL3 
is expected to be similar to that of the INPOL because their 
volume of the analyzer planes is similar with each other.
 In fact, the FOM value of $0.89\times 10^{-4}$ in Type-I 
is close to that of $0.81\times 10^{-4}$ of the INPOL.
 Thus we conclude that the performance of the NPOL3 is 
very similar to that of the INPOL by taking into account 
of the difference of the thicknesses of both analyzer and 
catcher planes.
 
\section{Summary}
\label{sec:summary}

 The high resolution neutron polarimeter NPOL3 has been 
constructed and developed to measure polarization transfer 
observables $D_{ij}$ for $(\vec{p},\vec{n})$ reactions 
at intermediate energies around $T_p$ = 200 MeV.
 The NPOL3 system can measure the normal as well as sideways 
components of the neutron polarization simultaneously.
 The other longitudinal component can be measured by using
the neutron spin rotation magnet described in 
Ref.~\cite{nima_369_120_1996}.
 Thus we can perform the measurement of a complete set of $D_{ij}$
with a high resolution of $\sim$300 keV by using the NPOL3.

 The performance of the NPOL3 system was studied by using 
the polarized neutrons from 
${}^{2}{\rm H}(p,n)pp$, 
${}^{6}{\rm Li}(p,n){}^{6}{\rm Be}(\mathrm{g.s.})$, and 
${}^{12}{\rm C}(p,n){}^{12}{\rm N}(\mathrm{g.s.})$ reactions at 
$T_p$ = 198 MeV.
 The effective analyzing powers in Type-I
are 0.33 and 0.12 for $(\vec{n},n)$ and $(\vec{n},p)$ channels,
respectively.
 The performance is comparable to that of the INPOL system
at the same energy.

\section{Acknowledgments}

 We are grateful to the RCNP Ring Cyclotron crew for their
efforts in providing a good quality beam.
 The experiments were performed at RCNP 
under program numbers E218 and E236.
 This work was supported in part by the 
Grant-in-Aid for Scientific Research No. 
14702005 of the 
Ministry of Education, Culture, Sports, 
Science, and Technology of Japan.


\clearpage

%
%

\clearpage

\begin{table}
\begin{center}
\begin{tabular}{ccccccc}
\hline
          &
\multicolumn{2}{c}{Type-I} &  
\multicolumn{2}{c}{Type-II} &  
\multicolumn{2}{c}{Type-III} \\
Quantity  &
\multicolumn{2}{c}{FOM optimum} &  
\multicolumn{2}{c}{$A_{y;\mathrm{eff}}$ optimum} &  
\multicolumn{2}{c}{Intermediate} \\
          &
$(\vec{n},n)$ & $(\vec{n},p)$ &
$(\vec{n},n)$ & $(\vec{n},p)$ &
$(\vec{n},n)$ & $(\vec{n},p)$ \\
\hline
$\Phi$ &
$67^{\circ}$ & $67^{\circ}$ & 
$67^{\circ}$ & $67^{\circ}$ & 
$67^{\circ}$ & $67^{\circ}$ \\
$\theta$ &
$( 8^{\circ},30^{\circ})$ &  $( 8^{\circ},53^{\circ})$ &  
$(12^{\circ},24^{\circ})$ &  $(20^{\circ},53^{\circ})$ &  
$(10^{\circ},27^{\circ})$ &  $(12^{\circ},53^{\circ})$ \\
$R_v$ &
$(0.90,1.30)$ &  $(0.70,1.30)$ &  
$(0.96,1.30)$ &  $(0.88,1.30)$ &  
$(0.92,1.30)$ &  $(0.70,1.30)$ \\
$\overline{\mathcal{L}}_{\mathrm{th}}$ &
$4\,\mathrm{MeV_{ee}}$ & $4\,\mathrm{MeV_{ee}}$ &
$4\,\mathrm{MeV_{ee}}$ & $4\,\mathrm{MeV_{ee}}$ &
$4\,\mathrm{MeV_{ee}}$ & $4\,\mathrm{MeV_{ee}}$ \\
\hline
\end{tabular}
\end{center}
\caption{Optimized cuts used in the NPOL3 system.}
\label{table:cut}
\end{table} 

\begin{table}
\begin{center}
\begin{tabular}{lrrrrrr}
\hline
         &
\multicolumn{2}{c}{Type-I} &  
\multicolumn{2}{c}{Type-II} &  
\multicolumn{2}{c}{Type-III} \\
Energy (Reaction)   &
\multicolumn{2}{c}{FOM optimum} &  
\multicolumn{2}{c}{$A_{y;\mathrm{eff}}$ optimum} &  
\multicolumn{2}{c}{Intermediate} \\
         &
$(\vec{n},n)$ & $(\vec{n},p)$ &
$(\vec{n},n)$ & $(\vec{n},p)$ &
$(\vec{n},n)$ & $(\vec{n},p)$ \\
\hline
193 MeV (${}^{2}{\rm H}(\vec{p},\vec{n})pp$) &
0.333 & 0.123 & 0.398 & 0.154 & 0.373 & 0.128 \\
                                            &
$\pm 0.008$ & $\pm 0.002$ & 
$\pm 0.013$ & $\pm 0.003$ & 
$\pm 0.009$ & $\pm 0.002$ \\
193 MeV (${}^{6}{\rm Li}(\vec{p},\vec{n}){}^{6}{\rm Be}(\mathrm{g.s.})$) &
0.305 & 0.123 & 0.369 & 0.148 & 0.341 & 0.136 \\
                                            &
$\pm 0.025$ & $\pm 0.009$ & 
$\pm 0.032$ & $\pm 0.011$ & 
$\pm 0.028$ & $\pm 0.010$ \\
180 MeV (${}^{12}{\rm C}(\vec{p},\vec{n}){}^{12}{\rm N}(\mathrm{g.s.})$) &
0.358 & 0.105 & 0.427 & 0.129 & 0.398 & 0.112 \\
                                            &
$\pm 0.021$ & $\pm 0.004$ & 
$\pm 0.034$ & $\pm 0.008$ & 
$\pm 0.024$ & $\pm 0.005$ \\
\hline
\end{tabular}
\end{center}
\caption{Effective analyzing powers of the NPOL3 system.}
\label{table:ayeff}
\end{table} 

\begin{table}
\begin{center}
\begin{tabular}{lcccccc}
\hline
         &
\multicolumn{2}{c}{Type-I} &  
\multicolumn{2}{c}{Type-II} &  
\multicolumn{2}{c}{Type-III} \\
Energy                   &
\multicolumn{2}{c}{FOM optimum} &  
\multicolumn{2}{c}{$A_{y;\mathrm{eff}}$ optimum} &  
\multicolumn{2}{c}{Intermediate} \\
         &
$(\vec{n},n)$ & $(\vec{n},p)$ &
$(\vec{n},n)$ & $(\vec{n},p)$ &
$(\vec{n},n)$ & $(\vec{n},p)$ \\
\hline
193 MeV &
$2.5\times 10^{-4}$ & $5.9\times 10^{-3}$ &
$0.9\times 10^{-4}$ & $1.7\times 10^{-3}$ &
$1.8\times 10^{-4}$ & $4.9\times 10^{-3}$ \\
180 MeV &
$2.8\times 10^{-4}$ & $6.5\times 10^{-3}$ &
$1.0\times 10^{-4}$ & $1.8\times 10^{-3}$ &
$2.0\times 10^{-4}$ & $5.4\times 10^{-3}$ \\
\hline
\end{tabular}
\end{center}
\caption{Double scattering efficiencies of the NPOL3 system.}
\label{table:eds}
\end{table} 

\end{document}